\documentclass{FM6_Focus_in_Astronomy}
\usepackage{graphicx}
\usepackage{textcomp}

\title[Kinematics of ETG stellar halos] 
{The extended Planetary Nebula Spectrograph (ePN.S) early type galaxy survey: the kinematic diversity of stellar halos}

\author[Claudia Pulsoni]   
{Claudia Pulsoni$^{1,2}$,
Ortwin Gerhard$^1$,
Magda Arnaboldi$^3$,
Lodovico Coccato$^3$,
\& PN.S Consortium}

\affiliation{$^1$Max-Planck-Institut f\"ur extraterrestrische Physik, \\ Giessenbachstra{\ss}e, 85748 Garching, Germany
\\[\affilskip]
$^2$ Excellence Cluster Universe, Boltzmannstra{\ss}e 2, 85748, Garching, Germany  \\[\affilskip]
$^3$ European Southern Observatory, Karl-Schwarzschild-Stra{\ss}e 2, 85748 Garching, Germany\\
 email: {\tt cpulsoni@mpe.mpg.de}}

\pubyear{...}
\jname{Astronomy in Focus, Volume 1} 
\editors{...}
\begin{document}

\maketitle

\begin{abstract}

In this contribution we report on a kinematic study for 33 early type galaxies (ETGs) into their outer halos (average 6 effective radii, Re).  We use planetary nebulae (PNe) as tracers of the main stellar population at large radii, where absorption line spectroscopy is no longer feasible.  The ePN.S survey is the largest survey to-date of ETG kinematics with PNe, based on data from the Planetary Nebula Spectrograph (PN.S), counter-dispersed imaging, and high-resolution PN spectroscopy.
We find that ETGs typically show a kinematic transition between inner regions and halos. Slow rotators have increased  rotational support at large radii. Most of the ePN.S fast rotators show a decrease in rotation, due to the fading of the stellar disk in the outer, more slowly rotating spheroid. 30\% of these fast rotators are dominated by rotation also at large radii, 40\% show kinematic twists or misalignments, indicating a transition from oblate to triaxial in the halo. Despite this variety of kinematic behaviors, the ePN.S ETG halos have similar angular momentum content, independently of fast/slow rotation of the central regions. 
Estimated kinematic transition radii in units of $R_e$ are $\sim1-3$ Re and anti-correlate with stellar mass. These results are consistent with cosmological simulations and support a two-phase formation scenario for ETGs.

\end{abstract}

\firstsection 
\section{Introduction}

Our understanding of the nature of early type galaxies (ETGs) has progressed remarkably in the last decade with the advent of the integral field spectroscopy (IFS). In the ATLAS3S survey (\cite[Cappellari et al. 2011]{2011MNRAS.413..813C}) ETGs have been found to divide among fast (FR) and slow rotators (SR) according to the angular momentum of their central regions (\cite[Emsellem et al. 2011]{2004MNRAS.352..721E}). The two classes have been interpreted as the result of different gas-rich or gas-poor merger processes that compete in the formation and evolution of these objects, while shaping their morphology and kinematics (\cite[Naab et al. 2014]{2014MNRAS.444.3357N}, \cite[Penoyre et al. 2017]{2017MNRAS.468.3883P}, \cite[Smethurst et al. 2018]{2018MNRAS.473.2679S}). Current surveys (MANGA, \cite[Bundy et al. 2015]{2015ApJ...798....7B}, CALIFA, \cite[S\'anchez et al. 2012]{2012A&A...538A...8S}, SAMI, \cite[Bryant et al. 2015]{2015MNRAS.447.2857B}, and MASSIVE, \cite[Ma et al. 2014]{2014ApJ...795..158M}) are working to increase the sample size of the IFS mapped objects, and extend the study to a wider range of environments and masses, but they are all focused to the central 1-2 effective radii ($R_e$). This limitation is naturally imposed by the low surface brightness in the outskirts of galaxies, which hampers spectroscopic measurements at larger radii. On the other hand the study of the kinematics in the outer regions is important since they contain half of the galaxies' stars and most of their dynamical mass. In addition, simulations suggest that the halos are mostly made of accreted material (\cite[Cooper et al. 2013]{2013MNRAS.434.3348C}, \cite[Rodriguez-Gomez et al. 2016]{2016MNRAS.458.2371R}, \cite[Pillepich et al. 2018]{2018MNRAS.475..648P}), hence the study of the halos implies exploring a different phase of the galaxies. 

The study of the kinematics at large radii requires alternative tracers that overcome the limit of low surface brightness in the outskirts. Planetary nebulae (PNe) are the optimal probes for the halo stellar kinematics. They are bright [OIII] emitters, which makes them easily detectable in narrow-band images over the faint halo. Because they are drawn from the main stellar population of the galaxy, their number density is proportional to the surface brightness, and their kinematics agrees with that from integrated light in the radial range of overlap: PNe follow light. This makes them reliable tracers for the kinematics of stars in the halos.

\section{ETG halo kinematics traced by PNe} 

The extended Planetary Nebula Spectrograph (ePN.S) survey of ETGs is based on data from the custom built Planetary Nebula Spectrograph at the William Hershel Telescope in La Palma, supplemented with catalogs of PN radial velocities from high-resolution spectroscopy and counter-dispersed imaging. 
With a total of 8636 PNe, the ePN.S survey is the largest kinematic survey to-date of extragalactic PNe in the outer halos of ETGs.
The sample of galaxies is magnitude limited, and contains 33 objects spanning a wide range of structural parameters, such as luminosity, ellipticity, central velocity dispersion (\cite[Arnaboldi et al. in prep.]{Arnaboldi_etal}). In this sample, 24 galaxies are FR and 9 are SR. 

In \cite[Pulsoni et al. (2018)]{Pulsoni_etal2018} we reconstructed halo velocity and velocity dispersion fields from the measured PN velocities using an adaptive kernel smoothing technique as in \cite[Coccato et al. 2009]{2009MNRAS.394.1249C}. This procedure is calibrated with simulations and finds the best compromise between spatial resolution and statistical noise smoothing. 
From the smoothed velocity fields we measured rotation velocity $\mathrm{V_{rot} (R)}$, kinematic position angle $\mathrm{PA_{kin}(R)}$, and velocity dispersion $\sigma(R)$ profiles.
The PNe extend the kinematic information out to a maximum radius of $[3-13]$ $R_e$, with a median of $5.6$ $R_e$ (average of $6 R_e$). The complement with absorption line data from the literature delivers a complete description of the kinematics from the central regions to the halos. 

We found that the halos of the ePN.S ETGs show a larger variety of kinematic behaviors in the halos than they do in the central regions. Most of the FR have a drop in $\mathrm{V_{rot}}$ in the halo. This property was already been observed by \cite[Coccato et al. (2009)]{2009MNRAS.394.1249C}, with PNe on a smaller sample of galaxies, and by \cite[Arnold et al. (2014)]{2014ApJ...791...80A}, using slitlets kinematics from the SLUGGS survey, who interpreted it as the fading of a rotating disk-like component into a more dispersion dominated spheroidal halo. In our sample we observed it in 70\% of the FR. 
While the central $\mathrm{PA_{kin}}$ of the ePN.S FR is well aligned with the photometric major axis (\cite[Krajnovi\'c et al. 2008, 2011]{2008MNRAS.390...93K, 2011MNRAS.414.2923K}, \cite[Foster et al. 2016]{2013MNRAS.435.3587F}), showing that the central regions are mostly axisymmetric, the measurement of the $\mathrm{PA_{kin}}$ at large radii reveals that the halo component might be triaxial for 40\% of these galaxies. 
The sample of SR shows increased rotation at large radii, and $\mathrm{PA_{kin}}$ generally twisting with radius or misaligned with the photometric position angle (${\mathrm{PA_{phot}}}$). This is, again, signature of a triaxial halo.
Most ePN.S galaxies have flat or slightly falling $\sigma(R)$ profiles, but 15\% of the sample have steeply falling profiles.

\section{Angular momentum at large radii}

\begin{figure}[b]
\begin{center}
 \includegraphics[width=0.49\columnwidth]{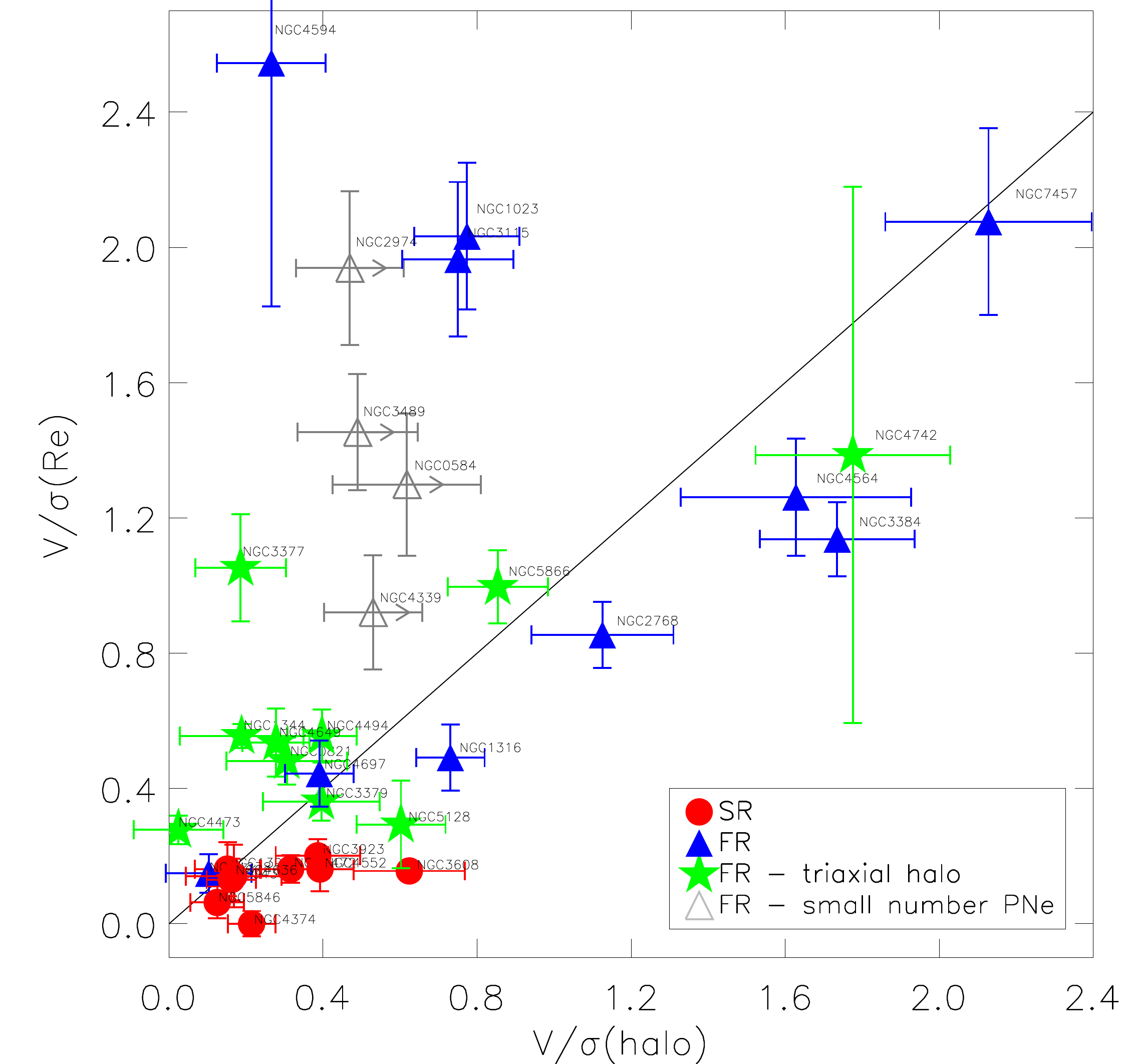} 
 \includegraphics[width=0.49\columnwidth]{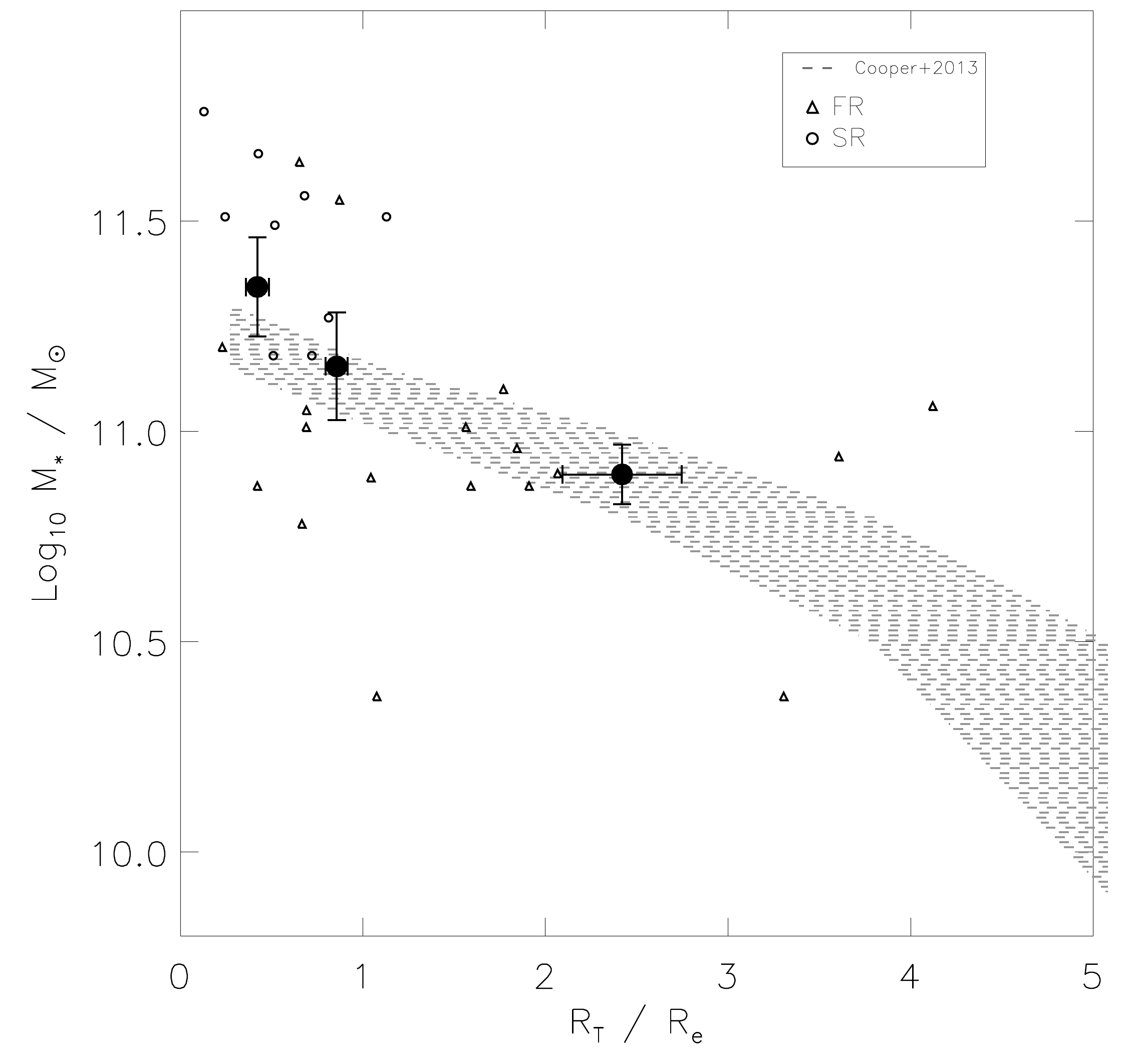} 
 \caption{Left panel: $\mathrm{V/\sigma(1R_e)}$ from absorption line data compared with $\mathrm{V/\sigma(halo)}$ from PN data. Different symbols show SRs, FRs, FRs with triaxial halos, and FRs with small numbers of PNe; for the latter $\mathrm{V/\sigma(halo)}$ is a lower limit estimate.
 Right panel: transition radius $R_T$ in units of $R_e$ as a function of stellar mass $M_{*}$. The shaded area shows the corresponding quantities from the simulations of \cite[Cooper et al. (2013)]{2013MNRAS.434.3348C}. Credit: Pulsoni et al., A\&A, 618, A94, 2018, reproduced with permission \textsuperscript{\tiny{\textcopyright}}ESO. }
   \label{fig1}
\end{center}
\end{figure}

Figure \ref{fig1} (left panel) compares the $V/\sigma$ ratio at $1 R_e$ and in the halo. The ePN.S SR crowd below the one-to-one line, showing the increased rotation in the halo.
Most of the ePN.S FR are still fast rotating in the halo, but have a lower $V/\sigma$ at large radii than in the central regions, comparable to the SR $\mathrm{V/\sigma(halo)}$. This uniformity in angular momentum content suggests the idea that the halos of these galaxies might actually be structurally similar, despite the differences in the central regions. 
A smaller group of ePN.S FR, representing 30\% of the sample, stands out for having particularly high $\mathrm{V/\sigma}$. These galaxies do not show any decrease in the $\mathrm{V_{rot}}$, and the high rotation may be due either to an extended disk-like component (NGC 7457), or to a rapidly rotating spheroid (NGC 2768).
The ePN.S FR showing kinematic signatures of a triaxial halo at the survey resolution span all values of $\mathrm{V/\sigma(halo)}$. 
The distribution of the $\mathrm{V/\sigma(halo)}$ values is reflected in the $\lambda_R(R)$ profiles in \cite[Coccato et al. in prep]{Coccato_etal}.

\section{Kinematic transition and dependence on stellar mass}

The variation of the kinematics from the central regions to the halos can be parametrized by defining a transition radius ($\mathrm{R_T}$), marking the radial range where the kinematics change significantly. In \cite[Pulsoni et al. (2018)]{Pulsoni_etal2018} we defined it using one of the following criteria:

\begin{itemize}
 \item in case of declining $\mathrm{V_{rot}(R)}$, $\mathrm{R_T}$ is the average of the radial interval between the maximum rotation $\mathrm{V_{rot}^{max}}$ and $\mathrm{V_{rot}^{max}} - 50$ km/s;
 \item in case of growing $\mathrm{V_{rot}(R)}$, $\mathrm{R_T}$ is the average of the radial interval between $\mathrm{V_{rot}} = 0$ km/s and $\mathrm{V_{rot}} = 50$ km/s;
 \item in case of kinematic twist, $\mathrm{R_T}$ is the average of the radial range in which $\mathrm{PA_{kin}(R)}$ changes significantly.
\end{itemize}

Figure \ref{fig1} (right panel) shows that $\mathrm{R_T}/R_e$ anti-correlate with stellar mass $M_{*}$, so that most massive galaxies have the transition in the kinematics at smaller fractions of $R_e$. A natural interpretation for the different kinematics of the halos with respect to the central regions is that halos have a different origin. The two phase formation scenario (e. g \cite[Trujillo et al. 2006]{2007MNRAS.382..109T} and \cite[Oser et al. 2010]{2010ApJ...725.2312O}) implies that galaxies are layered structures, in which the central regions are mostly formed by primordial (in-situ) stars and the halos are mainly made of accreted (ex-situ) stars (\cite[Cooper et al. 2013]{2013MNRAS.434.3348C}, \cite[Rodriguez-Gomez et al. 2016]{2016MNRAS.458.2371R}). More massive galaxies built up their mass by hierarchically accreting more satellites, and their ex-situ component dominates over the in-situ down to smaller radii. By comparison, lower mass galaxies have a lower accreted fraction, which is mainly deposited in the outskirts (\cite[Pillepich et al. 2018]{2018MNRAS.475..648P}). The shaded regions in Fig.\ref{fig1} report the transition radius between in-situ to ex-situ dominated regions as a function of $M_{*}$ from the particle tagging simulations of \cite[Cooper et al. (2013)]{2013MNRAS.434.3348C}. The agreement between the trend from the simulations and that from the observed kinematics supports the interpretation of $R_T$ as marking the transition between in-situ central regions and accreted halos. Hence the fact that the halos of the SRs and of the massive FRs in the ePN.S sample have similar angular momentum content can be explained by a common, ex-situ, nature.

\section{Summary}

PNe are reliable tracers for the kinematics of stars, hence they can be used to probe the outer regions of ETGs, which are too faint for absorption line measurements. In the ePN.S survey PNe extend the kinematic information out to $6 R_e$ on average. In the ePN.S sample ETGs show a larger diversity of kinematic properties in the halos than in their central regions. A small group of FRs, representing $\sim30\%$ of the FR sample stands out for having particularly high $V/\sigma$ ratio in the halo. On the other hand most of the galaxies show similar halo angular momentum, independently from their FR or SR status of their central regions. The fraction of high $V/\sigma\mathrm{(halo)}$ FRs might vary for samples of galaxies with different luminosity.
We defined a transition radius to describe the variation of the kinematics from the central regions to the halos, and found that this quantity anti-correlates with mass. This is consistent with cosmological simulations and supports a common ex-situ origin for the halos of these galaxies.

\end{document}